\documentclass{aa}
\usepackage{graphicx}
\usepackage{txfonts}
\usepackage[figuresright]{rotating}
\usepackage[]{natbib}
\bibpunct{(}{)}{;}{a}{}{,}
\begin{document}
\title{A Catalog of Edge-on Disk Galaxies
\thanks{
The full catalog tables, Table \ref{cat1} and Table \ref{cat2}, will be available in electronic form
at the CDS via anonymous ftp to cdsarc.u-strasbg.fr (130.79.128.5)
or via http://cdsweb.u-strasbg.fr/cgi-bin/qcat?J/A+A/ }}
\subtitle{From Galaxies with a Bulge to Superthin Galaxies}

   \author{Stefan J. Kautsch
          \inst{1}
	  \and
	  Eva K. Grebel
          \inst{1}
	  \and
	  Fabio D. Barazza
	  \inst{1,2}
          \and
          John S. Gallagher, III\inst{3}
	  }

   \offprints{S. J. Kautsch}

   \institute{Astronomical Institute, Department of Physics and Astronomy,
    University of Basel, Venusstrasse 7, CH-4102 Binningen, Switzerland\\
              \email{kautsch@astro.unibas.ch}\\
	 \and     
         Department of Astronomy, University of Texas at Austin, 1 University
	 Station C1400, Austin, TX 78712-0259, USA\\
	 \and
         Department of Astronomy, University of Wisconsin, 475 North Charter
	 Street, Madison, WI 53706-1582, USA\\
         }

   \date{Received 5 August 2005; accepted 2 September 2005}

   \abstract{Spiral galaxies range from bulge-dominated early-type galaxies
to late types with little or no bulge.  Cosmological models do not predict
the formation of disk-dominated, essentially bulgeless galaxies,
yet these objects exist.  A particularly striking and
poorly understood example of bulgeless galaxies are flat or superthin
galaxies with large axis ratios.  We therefore embarked on a study aimed at
a better understanding of these enigmatic objects, starting by compiling a
statistically meaningful sample with well-defined properties.  The disk
axis ratios can be most easily measured when galaxies are seen edge-on.  We
used data from the Sloan Digital Sky Survey (SDSS) in order to identify
edge-on galaxies with disks in a uniform, reproducible, automated fashion.
In the five-color photometric database of the SDSS Data Release 1 (2099
deg$^2$) we identified 3169 edge-on disk galaxies, which we subdivided into
disk galaxies with bulge, intermediate types, and simple disk galaxies
without any obvious bulge component.  We subdivided these types further
into subclasses: Sa(f), Sb(f), Sc(f), Scd(f), Sd(f), Irr(f), where the (f)
indicates that these galaxies are seen edge-on.  Here we present our
selection algorithm and the resulting catalogs of the 3169 edge-on disk
galaxies including the photometric, morphological, and structural
parameters of our targets.  A number of incompleteness effects affect our
catalog, but it contains almost a factor of four more bulgeless galaxies with 
prominent simple disks (flat galaxies) within the area covered here 
than previous optical catalogs, which were based on
the visual selection from photographic plates (cf.\ Karachentsev et al.\
1999).  We find that approximately 15\% of the edge-on disk galaxies in our
catalog are flat galaxies, demonstrating that these galaxies are fairly
common, especially among intermediate-mass star-forming galaxies.
Bulgeless disks account for roughly one third of our galaxies when also
puffy disks and edge-on irregulars are included.  Our catalog provides a
uniform database for a multitude of follow-up studies of bulgeless galaxies
in order to constrain their intrinsic and environmental properties and
their evolutionary status.

   }

\titlerunning{Edge-on Disk Galaxy Catalog}
\authorrunning{S. J. Kautsch et al.}

   \maketitle

\section{Introduction}
During the last decade an increasing number of studies of late-type edge-on
disk-dominated galaxies has been conducted, reflecting a growing interest
in understanding these galaxies in the framework of galaxy evolution and
cosmological models.  Models describing the chemodynamical evolution of
disk galaxies within a slowly growing dark matter halo can successfully
reproduce many of the observed properties of Milky-Way-type disk galaxies
\citep{saml03,saml04}.  Models with high merger rates as mandated in
hierarchical merger scenarios face a number of problems when comparing the
predicted properties of galactic subcomponents with observations \citep[e.g.,][]{abadi03}.  
It is even more difficult to succeed in producing
disk-dominated, essentially bulge-less late-type galaxies, making these
objects an evolutionary enigma.  In cold dark matter (CDM) simulations the
resulting disks are smaller, denser, and have lower angular momentum than
observed.  Major mergers increase the angular momentum \citep[e.g.,][]{gard01}, 
but also destroy disks, hence it seems unlikely that
simple disk galaxies suffered major mergers in the recent past.  Adding
feedback alleviates the angular momentum problem to some extent \citep[e.g.,][]{som03,robe05}. 
However, \citet{don04} point out that dark halos
that did not suffer major mergers have too low an angular momentum to begin
with. This prevents them from producing the observed extended disks
from the collapse of their associated baryons, since
the specific angular momentum of the gas cannot be {\em increased}\, by
feedback processes. 

Overall, disk galaxies show a multitude of different morphologies ranging
from disk galaxies with a substantial bulge and with high surface
brightness to bulgeless low-surface-brightness (LSB) galaxies \citep[e.g.,][]{schom92,mat99} 
and various complex bulge/disk combinations in between
\citep[e.g.,][]{mdg04}.  While certain properties such as the asymptotically
flat rotation curves seem to be shared by most disk galaxies, they differ
in other key properties such as surface brightness and scale length.   In
order to understand how these systems form and evolve we need to understand
the morphological systematics from bulgeless disks to disk galaxies with a
dominant spheroidal bulge. 

Historically, models consider disk formation as the result of the collapse
of a gaseous protogalaxy \citep[e.g.,][]{fall80}).  Disks may then form
from inside-out around a pre-existing classical bulge \citep{atha05} or via
smooth accretion of material \citep{stein02}.  On the other hand, intense
star formation will also lead to  the formation of a bulge or a dense
nucleus in a bulgeless disk galaxy, aided by the rapidity of the gas infall
and the total amount of the accreted material (either through infall or
mergers) \citep{nogu01}.  Also bars, formed via instabilities of a disk, can
transport material to the disk center.  The subsequent star formation may
build up an additional bulge component, which can then stabilize the disk
\citep{saml03}.

\citet{dalc97} propose a scenario where gas in low angular momentum
protogalaxies collapses efficiently, resulting in high-surface-brightness
galaxies.  Protogalaxies with high angular momentum and lower mass, on the
other hand, evolve into LSB galaxies.   \citet{dalc97} note that
``gravitational collapse in any hierarchical model with Gaussian initial
conditions leads to a broad distribution of halo masses and angular
momenta'', which could account for the observed range of properties.
\citet{dal04} found that galaxies with disk circular velocities $V_c >
120$ km s$^{-1}$ tend to show bulges. They suggest that these objects are
more gravitationally unstable, which can lead to fragmentation and
gravitational collapse along spiral arms and subsequently to smaller gas
scale heights, pronounced dust lanes, and star formation.  In this picture,
slowly rotating disks are stable and have low star formation rates,
implying also lower metallicities. 

While these scenarios offer a convincing and internally consistent
explanation for the nature of disk galaxies, the frequency and stability of
disk-dominated galaxies is surprising from the cosmological point of view.
Hierarchical models of galaxy formation include violent interaction phases
that should destroy disky systems \citep{stein03,tayl03}.  A better
knowledge of disk-dominated galaxies may hence be key for understanding
their formation, evolution, and survival.

The need for a homogeneous search for mainly bulgeless edge-on galaxies was
recognized by \citet{kar89}, who wanted to use these objects in order to
investigate large-scale streaming motions in the universe.  He used
photographic data in order to identify and catalog these systems.  The
resulting catalogs are the ``Flat Galaxy Catalogue'' (hereafter FGC)
\citep{kar93} and its extension, the ``Revised Flat Galaxy Catalogue''
(hereafter RFGC) by \citet{kar}.  This RFGC is an all-sky survey and
contains the largest published compilation of visually selected bulgeless
edge-on galaxies: 4236 objects in total.  A collection of edge-on disk
galaxies in the near infrared is provided in ``The 2MASS-selected Flat
Galaxy Catalog'' \citep{mit04}. Since the appearance of these highly
inclined disks is essentially needle-like and does not exhibit a distinct
bulge component \citet{kar89} called them ``flat galaxies''.  Flat galaxies
are thin edge-on spiral galaxies which seem to be (nearly) bulgeless and of
late morphological Hubble type (Sc/Sd and later).  A few years earlier,
\citet{goa79} and \citet{goa81} already called attention to edge-on
galaxies with extreme axial ratios.  They called these systems ``superthin
galaxies''. Superthin and flat galaxies belong to the same group, which we
will summarize here under the term ``simple disk galaxies''. 

In order to contribute to a better characterization of these objects, we
carried out the work presented here, which aims at compiling an uniform
sample of disk-dominated galaxies from modern CCD data at optical
wavelengths.  The Sloan Digital Sky Survey (SDSS) with its homogeneous,
deep, large-area coverage provides an ideal data base for the
identification of such galaxies.  The SDSS \citep{yor00} is carrying out
multi-color imaging of one quarter of the sky, followed by
medium-resolution spectroscopy primarily of galaxies and other objects of
interest down to certain magnitude limits.  These data are pipeline-reduced
and the resulting images, astrometry, photometry, structural parameters,
and calibrated spectra are released to the public after a proprietary
period \citep{sto02,aba03,aba04a,aba04b}.

The SDSS with its resolution, dynamic range, and photometric accuracy
allows one to study statistical properties and biases of disk galaxies such
as their structure, intrinsic properties, overall frequency, and global
scaling relations.  The formation and evolution scenarios can be probed by
studying the detailed structure and morphology (e.g., bulges, bars, halos, 
knots, and lanes) and comparing these with predictions from models
\citep[e.g.,][]{saml03,saml04,Immel04}.  
Also the frequency of warps of the edge-on galaxies and possible relations with the 
environment can be studied easily using the SDSS. 
Warps should be relativy frequent since \citet{resh95} showed that about 40\% of the FGC
galaxies have pronounced warps.
Radial and vertical color gradients
in these systems can shed light on the assembly of structure and on the
evolutionary state using the available multi-color photometry.  

In addition, the SDSS spectra enable
the estimation of the properties of the stellar populations, of the  star
formation rates, central activity, and metallicities. The redshifts from
the catalog allow one to estimate the luminosities and sizes of the
galaxies and the  distribution of these properties. Also the environment of
the cataloged galaxies can be investigated to probe the distribution of
the surrounding satellites, the Holmberg effect, external influences on
morphological evolution, and the local density and properties (frequency,
position, and alignment) in a cluster environment.

We confined our search of the SDSS data base to edge-on disk galaxies,
which facilitates the definition of an effective selection criterion.  The
choice of edge-on galaxies in particular is the only way to reliably select
pure disk galaxies based on their optical morphologies.  
Altogether, we
collected 3169 edge-on galaxies from the SDSS Data Release 1 (SDSS DR1). 
These systems
can be subdivided into subclasses according their appearance. 

This paper is organized as follows: In \S  \ref{DA} we describe the
training set of galaxies and the resulting selection criteria.  In \S 3 the
actual target selection is explained.  The classification of the detected
objects is presented in \S 4, followed by a description of the catalog (\S
\ref{TC}).  Comparisons to evaluate the completeness of our selection are
presented in \S 6.  The influence of dust extinction on the galaxy
selection is discussed in \S 7.  Different types of galaxies and their
subclasses are discussed in \S \ref{Di}. The last Section \S \ref{CS}
contains the summary and conclusions.

\section{Training set and selection criteria} \label{DA}

\subsection{The data base}

Our intent is to find edge-on galaxies with dominant stellar disks.   We
are using Karachentsev et al.'s catalogs as a starting point in order to
carry out a systematic and reproducible selection of these kinds of galaxies. 
The object
selection in the FGC and RFGC was based on the visual identification of
galaxies with an axial ratio of a/b $\geqslant$ 7 and a major axis diameter
a $\gtrsim$ $40''$ in the blue band on copies of the POSS-I and ESO/SERC photographic
plates.  The availability of the SDSS database permits us to carry out a
survey using deep, homogeneous, five-color CCD data that are superior to
the less deep, inhomogeneous photographic plates.  An added advantage of
the SDSS is that it will ultimately allow us to carry out such a search in
an automated, objective, repeatable fashion.  This certainly does not
render the earlier studies superfluous since the SDSS is not an all-sky
survey and since the earlier identifications provide a valuable training
set for the definition of the selection criteria to be applied to the
digital data.  Furthermore, the SDSS permits us to identify not only
simple-disk candidates, but also edge-on galaxies in general and to
investigate the properties of all of these different morphological types.

We have analyzed SDSS data from DR1 \citep{aba03}, which was the largest
publicly available data set when this work was started.  DR1 provides 2099
deg$^2$ of imaging data observed in the five SDSS filters {\em ugriz}.  The
$r$-band depth of these data is approximately 22.6 mag.  Meanwhile the data
releases 2 (DR2) \citep{aba04a}, 3 (DR3) \citep{aba04b}, and 4 (DR4) \citep{adel05} are
available, which cover successively larger areas on the sky.  As detailed
in \citet{aba04a}, changes were made to the data processing software
between DR1 and DR2, but no such changes occurred for DR3 as compared to
DR2 \citep{aba04b}.  We compared the SDSS photometry parameters in DR1 and
DR2 for our galaxies and found no significant changes.  However, in all of
the releases some galaxies are affected by so-called ``shredding''
\citep[e.g.,][]{aba03,aba04a,kniazev04a}, i.e., these galaxies are detected
as two or more independent objects.  This is found in particular for
extended objects with substructure and diameters $\geqslant 1'$.  A
comparison of the different data releases showed that shredded target galaxies are
similarly miss-classified in all of these releases.  Some of the galaxies
that were correctly identified in DR1 turned out to be shredded in the
later releases.  Hence we decided to continue to work with DR1 for the
pilot study presented here. 

\subsection{Definition of a training set}

In order to quantify a training set for the selection of disk-dominated
edge-on galaxies we searched for all RFGC galaxies with a right ascension
between 00 00 00 and 02 12 00 in the DR1 database, using the RFGC
coordinates.  In this coordinate range we expect to recover 47 RFGC
galaxies in the DR1.  It turned out that two of these objects have 
significantly different  
coordinates from the galaxies detected in the SDSS, while a third
galaxy has a very different angular diameter in the RFGC as compared to the
SDSS.  For the remainder, the difference between the RFGC coordinates and
the SDSS coordinates is typically smaller than $\pm$0.001 degrees
($3.6''$).  For this ``training set'' that we re-identified in the SDSS, we
found that the structural parameters have slightly smaller values in the
DR1 as compared to the RFGC.  We tested various combinations of SDSS
structural and photometric parameters that would allow us to recover the
galaxies in the training set (and additionally other edge-on disk galaxies
in the SDSS).  Ultimately, these galaxies should be recovered by performing
an automated search of the SDSS photometric catalog database.

\subsection{Definition of the query}

As the result of this empirical approach, we finally adopted the parameters
listed below for subsequent queries of the DR1 ``Best Galaxy Table''
\citep{aba03}.  The DR1 ``Best Galaxy Table'' is the table in the SDSS
database containing all parameters for galaxies that are of the
highest quality at the time of the data release.
\begin{itemize}
\item Axial ratio in the g band: $a/b > 3$, where $a$ and $b$ are 
the major and minor axis, respectively.
\item Angular diameter (isophotal major axis of the galaxy in the
``blue'' g filter) $a > 30''$.
\item Colors in the range of $0.5 < g-r < 2$ mag and  
$0.5 < r-i < 2$ mag. 
\item Magnitude limit in the g filter $< 20$ mag. 
\end{itemize}

With these conditions we are able to essentially reproduce all of the RFGC
criteria and to recover the training set. In addition to the RFGC galaxies,
our query parameters yield a much larger number of flat edge-on galaxies
and other disk-dominated objects.  This increase in numbers is in part due
to the higher resolution and depth of the DR1 as compared to the
photographic plates, but also due to our intention to collect all edge-on
disk galaxies (including those with bulges).  The latter is facilitated
particularly by our relaxed choice of minimum axis ratios.  

The images of the thus found objects were then downloaded from the DR1
``Data Archive Server'' (DAS) using the SDSS rsync server.  We downloaded
the so-called ``corrected imaging frames'' (fpC) in the five SDSS bands.
For a detailed description of the fpC frames we refer to the on-line
description in the SDSS
webpages\footnote{http://www.sdss.org/dr1/dm/flatFiles/fpC.html}.

\section{Target selection} \label{os}

After a visual inspection we removed contaminants from our object list. The
contaminants are mostly spikes (from very bright stars) and artifacts such
as satellite or meteorite tracks which resemble an edge-on galaxy. Also
strongly spike-blended edge-on galaxies were rejected. Additionally,
obvious non-edge-on systems and unknown objects were removed. The obvious
non-edge-on systems are objects where a bright bar in a faint disk
simulates an edge-on disk. Apart from these contaminants, our selection
criteria produce a fairly uniform sample of extended disk-dominated
galaxies including objects with small bulges and bulgeless simple disks.
All in all, 3169 objects were assembled in our catalog.  Some early-type
edge-on spiral galaxies are also included in the catalog, but internal dust
lanes introduce a bias in excluding a fraction of these galaxies.  This
will be discussed in greater detail in Section \ref{dust}.  Additionally,
our sample is limited by our selection criteria and by the SDSS photometry
itself. The following biases affect our selection: (1) Edge-on galaxies
with very faint disks around bulges and bright centers.  (2) ``Shredded
galaxies''.  (3) Galaxies with unusual colors caused by an AGN and/or dust.

The edge-on galaxies remaining in our sample after visual inspection
and removal of contaminants fall into three general morphological
groups:
\begin{itemize}
\item Pure bulgeless disks/simple disks
\item Galaxies with a disk and an apparent bulge
\item Objects with disks and central light concentration but no 
obvious bulge-like structure.  These may be considered 
an intermediate class between the simple disks and galaxies with bulges.
This group also comprises edge-on disky irregulars.
\end{itemize}

Out of these galaxies, an effort was made to select by eye objects spanning
the full range in properties including different disk thicknesses,
different bulge sizes, and presence or absence of dust lanes. The result is
a subsample of 129 galaxies that is our morphological ``reference set''.
Via visual inspection we subdivided this reference set into 42 simple
disks, 37 galaxies with a bulge, and 50 intermediate types with central
light excess.  We then used this subsample to further automate the
separation process and to develop a code to recover these general classes
of edge-on galaxies in the SDSS DR1. 

We found that the luminosity-weighted mean value of the ellipticity
(hereafter $\varepsilon$) of the elliptical isophotes is a very robust
separator between simple disks and the other edge-on types.  In combination
with the concentration index (hereafter CI) we can also exclude galaxies
with an apparent bulge. The CI clearly separates galaxies with bulge from
those without an apparent spheroidal component.  This will be detailed in
the following sections.

\subsection{Isophote fitting}

The following analysis is performed with the MIDAS analysis package
developed by the European Southern Observatory.  We applied it to the
frames in all five SDSS filters, but only used the results for the three
most sensitive bands {\em gri}.  Unless explicitly specified otherwise,
magnitudes quoted below refer to each of the separate bands.  Firstly, we
subtract the sky and the ``softbias'' from all frames.  The softbias is an
additional offset of 1000 counts per pixel in order to avoid negative
pixels in the images. The sky and softbias were subtracted as mean values
from the images.  The values of the sky and softbias are stored in the
header of each fpC frame. Then we use the MIDAS surface photometry package
``surfphot'' to fit ellipses to the isophotes of our galaxies. The
innermost ellipse is fitted adopting the center coordinates given by the
DR1 photometric database.  The intensity of the innermost isophote is
derived from the luminosity of the brightest pixel in a box ($8 \times 8$
pixel) that corresponds to the galaxy center. 

In steps of 0.2 mag the program fits ellipses until an isophote is
reached that corresponds to a surface brightness of $\mu = 25$ mag
arcsec$^{-2}$.  This implies that on average 20 -- 30 isophote levels are
plotted for every galaxy depending on the size and brightness.  This
isophote algorithm is based on the formulae of \citet{ben87}.

\subsection{Measuring the luminosity-weighted mean ellipticity and 
concentration index}

We use the resulting values of the isophote levels and the major (a)
and minor (b) axes in order to derive the luminosity weighted mean
ellipticity of the elliptical isophotes ($\varepsilon$). $\varepsilon$
is defined as 

\begin{equation}
\varepsilon = \frac{\sum \limits_{i=1}^{n} \epsilon_i \cdot I_i}{\sum 
\limits_{i=1}^{n} I_i}
\end{equation}
and 
\begin{equation}
\epsilon_i = 1 - \frac{b_i}{a_i}
\end{equation}
is the ellipticity of the {\em i}th isophote, whereas
\begin{equation}
I_i = z_i \cdot ((a \cdot b)_i - (a \cdot b)_{i-1}) \cdot \pi
\end{equation}
is the intensity between two isophote levels. The isophote level is 
indicated by $z_{i}$.

For the CI of these objects we
used the ratio of the following SDSS parameters 
\begin{equation}
CI = petrorad\_90 \cdot petrorad\_50^{-1}
\end{equation}
This is the ratio of the Petrosian radii (petrorad) that contain 
90\% and 50\% of the Petrosian flux in the same band, respectively 
\citep[see][]{sto02}.  
The Petrosian radius 
is the radius of a circular
aperture at which the ``Petrosian ratio'' is set to a fixed value of 0.2.
This
``Petrosian ratio'' is the ratio of the surface brightness in an
annulus at a certain radius to the mean surface brightness within a
circle with this radius. As discussed in \citet{strauss02} 
the use of circular apertures instead of elliptical apertures is
fairly insensitive to inclination.  Similarly,
the Petrosian magnitudes are derived from the
Petrosian flux using a circular aperture centered on every object. The advantage
of this method is that this allows an unbiased measurement of a constant
fraction of the total galaxy light using the technique based on that of
\citet{petro76}. For a detailed description of the Petrosian parameters used in
the SDSS we refer to \citet{blan01} and \citet{yasu01}.

The CI is known as a morphological separator between early- and late-type
galaxies \citep[see, e.g.,][]{str01,shi01,nak03,she03,kniazev04a}.  With the
CI and with $\varepsilon$ as separators we recover our visually selected
subgroups of the training set very well automatically. Therefore we applied
this procedure to all edge-on galaxies in our catalog.

\subsection{Choosing the limiting values of CI and $\varepsilon$}

\begin{figure}
\centering
\resizebox{\hsize}{!}{\includegraphics[angle=0,width=\textwidth,]{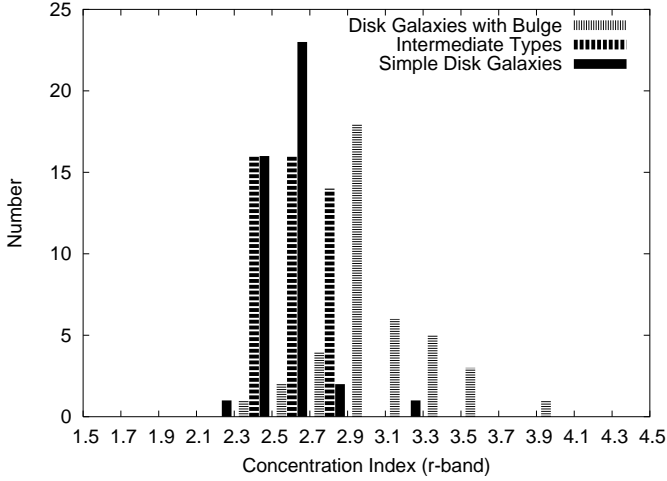}}
\caption{Number distribution of the visually selected galaxies versus their concentration index (CI).}
\label{ci}
\end{figure}

\begin{figure}
\resizebox{\hsize}{!}{\includegraphics[angle=0,width=\textwidth]{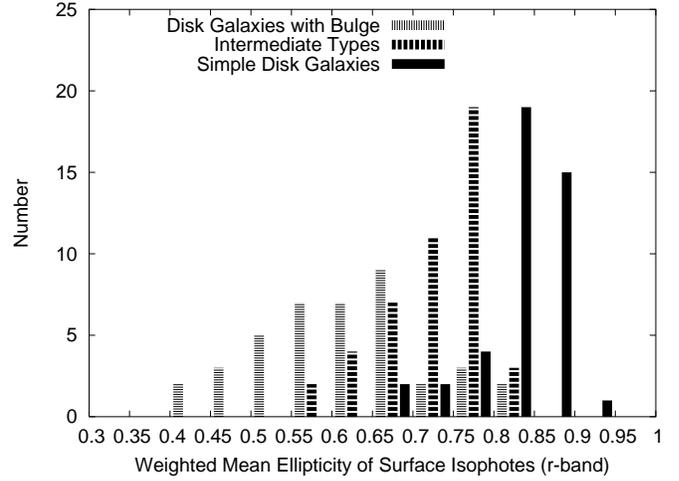}}
\caption{Number distribution of the visually selected galaxies versus their 
weighted mean ellipticity of the isophotes ($\varepsilon$).}
\label{e}
\end{figure}

\begin{figure}
\resizebox{\hsize}{!}{\includegraphics[angle=0,width=\textwidth]{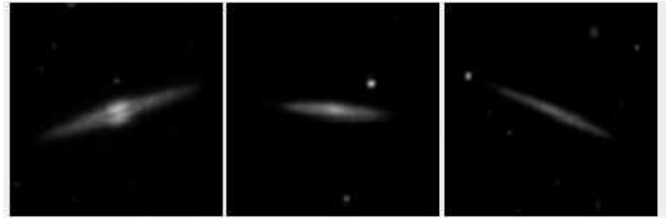}}
\caption{The left image is a typical
member of the class of galaxies with bulge Sb(f): \object{SDSS J020405.91-080730.3}. 
A typical example of the Scd(f) intermediate class
is in the
middle: \object{SDSS J102903.90+611525.8}. Simple disk galaxies Sd(f) have an appearance like 
\object{SDSS J135309.65+045739} at the right. All images are cutouts from the DR3 Image List Tool. These images have
a scale of 90 square arcsec. }
\label{exa}
\end{figure}

In order to determine which choices of CI  distinguish best between the
general types of edge-on galaxies we use a histogram with the distribution
of the CIs of the visually ``classified'' galaxies in our training
subsample (Figure \ref{ci}). Clearly, the majority of the simple disk
galaxies has a CI $< 2.7$, hence we adopt a CI of 2.7 as the boundary
condition to differentiate between simple disks and intermediate-type
galaxies from those of with an obvious bulge component.  A slightly lower
value was often used in previous morphological studies of galaxies from the
SDSS in order to separate between S0/Sa-type spirals and later spiral types
\citep{str01,shi01,nak03,she03}.  Contrary to our study these authors did
not limit their samples to edge-on galaxies.  Disky irregular galaxies
exhibit very low CIs: the limit is CI $< 2.15$ \citep[see also discussion
in][]{kniazev04a}.  Unfortunately it is not possible to separate the
intermediate type from simple disks using only the CI. As one can see in
the histogram the two remaining classes are merged at low values of the CI
despite their different morphological appearance.

\begin{figure*}
\includegraphics[angle=0,width=\textwidth]{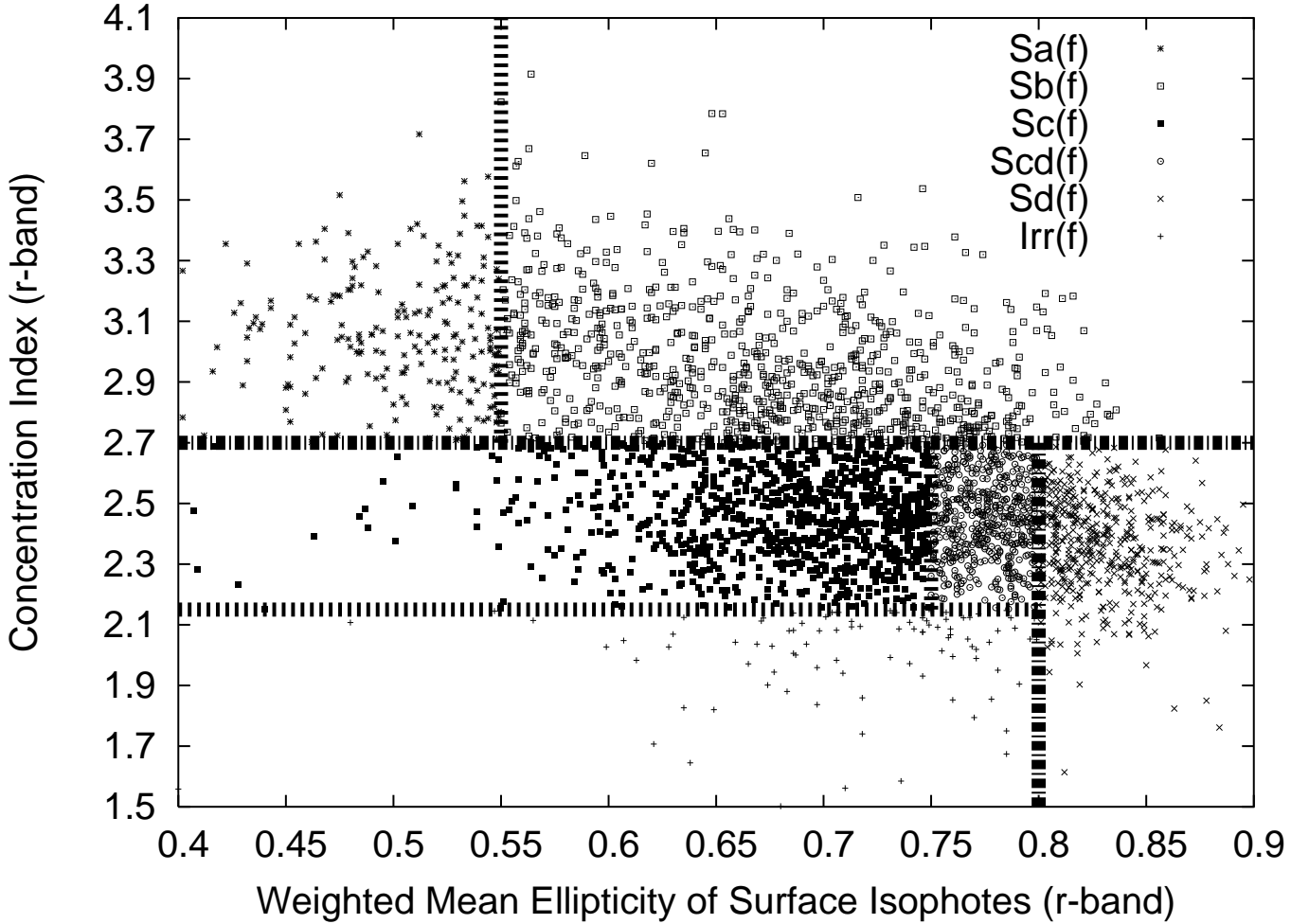}
\caption{The main separation diagram. The symbols are cited in the key and represents the various 
morphological types. The borders of the general classes are marked with long-short dotted lines. The additional 
short dotted lines
corresponds to the subgroup boarders within the general classes.}
\label{sd}
\end{figure*}

For that reason we use the weighted mean ellipticity of the isophotes
$\varepsilon$ as the second discriminator. Other possible morphological
separators from the literature such as colors, asymmetry index, and profile
likelihoods \citep{str01}, as well as decomposition \citep{k4} and Gini
index \citep{abr03,lot04} turned out not to be useful for the sensitive
characterization of edge-on galaxies \citep{kg3} probably because of the
influence of dust and of the galaxy inclination on these separators.

We again use a histogram of the number distribution (Fig.\ref{e}) of the
$\varepsilon$ values of the training subsample. In this case we intend to
separate the intermediate types from the simple disks. We defined the
region of 0.75 $\leq$ $\varepsilon$ $<$ 0.8 as the transition zone between
the classes and a value of $\varepsilon =$ 0.8 as the sharp border. This
limiting value allows us to select the best simple disk candidates and the
transition types. Additionally, with $\varepsilon < 0.55$ we can divide the class of 
galaxies with a bulge into early and later types.

\section{The classification of edge-on galaxies}

In order to flag these systems we follow the terminology introduced by
\citet{devo59}. In his scheme, spiral galaxies are marked with an
additional letter referring to the shape of the spiral arms, e.g., ``r''
means ring shaped and ``s'' s-shaped spiral structure when seen face on. We
will instead use an ``f'' to indicate that a galaxy is flat, i.e., contains
an edge-on component with or without a bulge.  Furthermore, we introduce
the following subclasses:  galaxies with bulges (Sa(f), Sb(f)); simple
disks (Sd(f)), Sc(f) and an intermediate group between Sc(f) and Sd(f)
called Scd(f), and disky edge-on irregulars (Irr(f)). 
Representative examples of the general class members are shown in Fig.\ref{exa}.  
The three galaxies shown are for the Sb(f) class (\object{SDSS J020405.91-080730.3}), 
for the Scd(f) class (\object{SDSS J102903.90+611525.8}), and
for the Sd(f) class (\object{SDSS J135309.65+045739.3}).  These galaxy images are
three-color ({\em g, r, i}) composites provided by the SDSS DR3 Image List
Tool and have a scale of $90''$ in X and Y direction.
The separation diagram, Figure \ref{sd}, exhibits CI and
$\varepsilon$ in order to separate these classes.  The abscissa represents the
luminosity-weighted mean ellipticity of the isophotes $\varepsilon$. The
ordinate shows the concentration index CI as taken from the SDSS. The
values are given in the SDSS r band\footnote{ For the selection from the
photometric database we used the g band in order to make the parameters
comparable to those used for selecting the FGC. In the following diagrams,
however, we refer to the r band.  This filter is mostly used in the other
studies involving the CI (\citet{shi01,nak03,she03}) because its quantum
efficiency is the highest of all SDSS bands \citep{sto02}. In addition it
includes the red light of the bulge which is important to separate galaxies
with bulges from bulgeless galaxies.}.  The automatically recovered simple
disk galaxies are hereafter marked with Sd(f), the intermediate types with
Scd(f) and Sc(f), the galaxies with bulges with Sa(f) and Sb(f), and irregulars with
Irr(f).  The lines mark the borders between the general classes and the
subtypes in our new classification within the general classes.
The selection
parameters for the {\em g, r, i} bands are listed in Table \ref{tbl-1}.

\begin{figure}
\resizebox{\hsize}{!}{\includegraphics[angle=-90,width=\textwidth,]{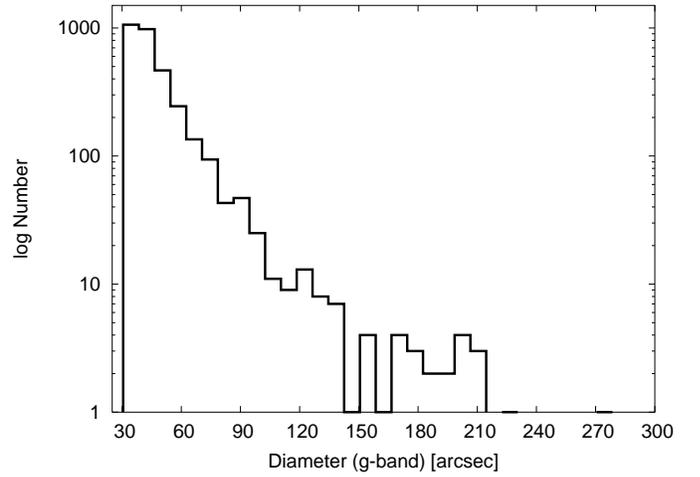}}
\caption{Logarithmic number distribution of the angular diameters of the catalog galaxies.}
\label{dia}
\end{figure}

\begin{figure}
\resizebox{\hsize}{!}{\includegraphics[angle=0,width=\textwidth,]{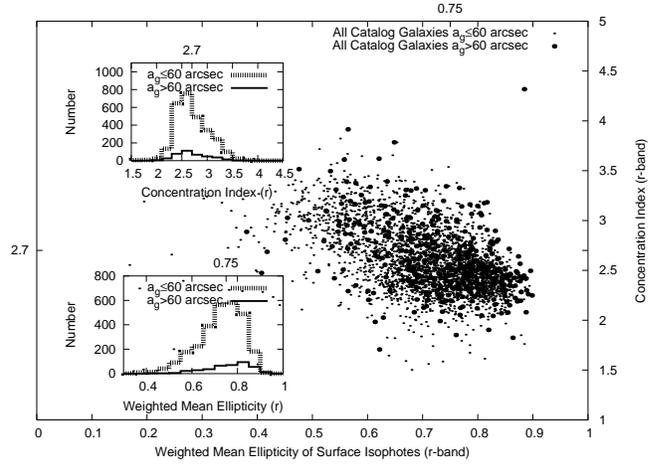}}
\caption{Separation diagram with the emphasized size samples. Galaxies with
angular diameters larger than 60 arcsec are indicated with filled points,
smaller galaxies with fine dots. The upper inset shows the number
distribution of the concentration index, the bottom inset that of the
weighted mean ellipticity. Objects with a diameter $a \leqslant$ 60 arcsec
are denoted by the dashed line. The others are indicated by a filled black
line.}
\label{sddia}
\end{figure}

In the histogram in Fig.\ \ref{dia} we plotted the number distribution of
the apparent diameters of our galaxies.  The majority of our objects is
smaller than $60''$ ($\sim$ 88\% of all catalog objects).  Only $\sim$ 2\%
of all galaxies in our catalog have a diameter larger than $100''$ (Fig.\
\ref{dia}). In order to check the influence of the size of the objects on
our separation we plotted galaxy samples with different angular size (less
than and greater than $60''$) in Fig.\ \ref{sddia}.  This permits us to
test whether higher resolution affects the separation process, in
particular whether bright centers and extended disks bias the
classification.  The upper inset shows the number distribution of the CIs.
Both size samples follow the same distribution.  In the bottom inset the
number distribution of $\varepsilon$ is presented. 

\begin{table}
\caption{Limiting Values. These are the values of the limits of the
morphological classes.  The values are valid for the SDSS g and r bands.
The value for the i band is the same as in the other filters for
$\varepsilon$.   For CI it is slightly higher because i is more sensitive for
the dominant redder bulge stars. In this case one should add a value of 0.1
to the numbers of the CI in this table. In general, note that the galaxies
near boundaries have the least certain classification.}
\label{tbl-1}
\centering
\begin{tabular}{rrr|rr}
\hline\hline
Class &  \multicolumn{2}{c}{$\varepsilon$} & \multicolumn{2}{c}{CI} \\
\cline{2-3} \cline{4-5} \\
  & lower limit & upper limit &  lower limit & upper limit\\
\hline
Sa(f) & $ -- $ & $< 0.55 $ & $\geq 2.70 $ & $ -- $ \\
Sb(f) & $\geq 0.55 $ & $ -- $ & $\geq 2.70 $ & $ -- $ \\
Sc(f) & $ -- $ & $< 0.75 $ & $\geq 2.15 $ & $< 2.70 $  \\
Scd(f) & $\geq 0.75 $ & $< 0.80 $ & $\geq 2.15 $ & $< 2.70 $ \\
Sd(f) & $\geq 0.80 $ & $ -- $ & $ -- $ & $< 2.70 $ \\
Irr(f) & $ -- $ & $< 0.80 $ & $ -- $ & $< 2.15 $ \\

\hline
\end{tabular}
\end{table}
Galaxies with diameters $a > 60''$ tend to have slightly higher values of
$\varepsilon$, since these galaxies tend to be closer to us, facilitating
the detection of more highly eccentric isophotes in the outer regions of
these extended objects.  

Using visual inspection we found that all Sd(f) types with an angular
diameter $a > 60''$ show the appearance of a simple disk. Consequently they
are assigned to the correct class by the automated algorithm. We therefore
conclude that the defined limiting values of our catalog are still robust
enough so that size and resolution do not affect the classification. \
The influence of resolution on the separation is discussed in section \ref{dust}.

\section{The Catalog} \label{TC}

\begin{sidewaystable*}
\caption{Catalog: Structural Parameters. The complete version of this 
table is available from the CDS.
Here we are presenting the first 15 entries (out of 3169) as an example.}
\label{cat1}
\centering
\begin{tabular}{lrrrllllllllllll} 
\hline\hline             
Name  & RA (J2000)  & DEC (J2000)  &Class  & $\varepsilon$(g) & $\varepsilon$(r) & $\varepsilon$(i) &CI(g) &CI(r)
 & CI(i)  & a(g)& a(r)& a(i) & a/b(g)  & a/b(r) & a/b(i)\\
\hline
\object{SDSS J000003.33-104315.8} & 00 00 03.335  & -10 43 15.899 & Sb(f)  &  0.746 &  0.740 &  0.735 &  2.801 &  3.032 &  3.162 &   37.730 &   41.511 &   42.443 &    5.863 &    5.294 &    4.469  \\  
\object{SDSS J000151.29-092430.3} & 00 01 51.294  & -09 24 30.368 & Scd(f) &  0.815 &  0.785 &  0.779 &  2.091 &  2.286 &  2.337 &   32.166 &   31.807 &   32.652 &    5.373 &    4.396 &    4.453  \\  
\object{SDSS J000222.39-102543.6} & 00 02 22.395  & -10 25 43.635 & Sd(f)  &  0.854 &  0.838 &  0.827 &  2.046 &  2.286 &  2.289 &   32.657 &   24.546 &   25.162 &    9.395 &    4.871 &    4.213  \\  
\object{SDSS J000301.47-001901.8} & 00 03 01.470  & -00 19 01.862 & Sb(f)  &  0.625 &  0.635 &  0.629 &  3.031 &  3.406 &  3.556 &   32.782 &   33.914 &   35.620 &    5.102 &    3.859 &    3.361  \\  
\object{SDSS J000347.01-000350.3} & 00 03 47.017  & -00 03 50.305 & Sc(f)  &  0.749 &  0.713 &  0.691 &  2.497 &  2.381 &  2.750 &   35.999 &   34.902 &   37.593 &    5.507 &    4.446 &    4.339  \\  
\object{SDSS J000530.43-095701.2} & 00 05 30.432  & -09 57 01.224 & Sd(f)  &  0.870 &  0.828 &  0.811 &  2.132 &  2.246 &  2.307 &   37.430 &   34.892 &   37.712 &    5.941 &    4.732 &    4.915  \\  
\object{SDSS J000542.53-111048.8} & 00 05 42.531  & -11 10 48.860 & Sb(f)  &  0.770 &  0.767 &  0.761 &  2.868 &  2.754 &  2.922 &   41.873 &   36.395 &   43.693 &    5.972 &    4.536 &    5.789  \\  
\object{SDSS J000619.88-092744.8} & 00 06 19.884  & -09 27 44.894 & Sc(f)  &  0.703 &  0.694 &  0.688 &  2.478 &  2.547 &  2.549 &   33.052 &   35.984 &   39.172 &    3.804 &    3.665 &    3.728  \\  
\object{SDSS J000628.86-004702.9} & 00 06 28.865  & -00 47 02.918 & Sa(f)  &  0.505 &  0.527 &  0.519 &  2.845 &  2.874 &  2.972 &   45.955 &   55.102 &   58.408 &    4.187 &    4.241 &    3.734  \\  
\object{SDSS J000641.10-105825.9} & 00 06 41.101  & -10 58 25.946 & Sc(f)  &  0.696 &  0.705 &  0.651 &  2.539 &  2.652 &  2.669 &   39.452 &   44.229 &   47.994 &    3.658 &    3.574 &    3.212  \\  
\object{SDSS J000741.22-004145.3} & 00 07 41.222  & -00 41 45.350 & Sc(f)  &  0.675 &  0.667 &  0.650 &  2.481 &  2.699 &  2.715 &   31.690 &   35.123 &   32.730 &    3.758 &    3.249 &    2.406  \\  
\object{SDSS J000919.54+003557.2} & 00 09 19.547  & +00 35 57.258 & Sb(f)  &  0.718 &  0.683 &  0.677 &  2.973 &  3.204 &  3.302 &   51.629 &   53.034 &   53.238 &    6.022 &    4.868 &    4.462  \\  
\object{SDSS J000924.17+003216.2} & 00 09 24.172  & +00 32 16.245 & Scd(f) &  0.810 &  0.795 &  0.814 &  2.227 &  2.506 &  2.136 &   30.648 &   32.993 &   32.352 &    6.731 &    6.445 &    5.573  \\  
\object{SDSS J000941.16-003152.4} & 00 09 41.166  & -00 31 52.414 & Sa(f)  &  0.498 &  0.502 &  0.513 &  3.026 &  3.051 &  3.129 &   40.068 &   42.329 &   43.317 &    4.367 &    3.656 &    3.231  \\  
\object{SDSS J000949.38-004103.1} & 00 09 49.382  & -00 41 03.194 & Scd(f) &  0.791 &  0.782 &  0.767 &  2.529 &  2.586 &  2.599 &   33.298 &   31.220 &   31.124 &    6.737 &    5.007 &    4.007  \\  
\hline
\end{tabular}
\end{sidewaystable*}

\begin{sidewaystable*}
\caption{ Photometric Parameters and Redshifts. The complete version 
of this table is available from the CDS.
Here we are presenting the first 15 entries (out of 3169) as an example.}
\label{cat2}
\centering
\begin{tabular}{lrrllllllllllll} 
\hline\hline             
Name & Class & mag(g) & mag err(g) & mag(r) & mag err(r) & mag(i) & mag err(i)& $\mu$(g) & $\mu$(r) &
$\mu$(i)  & z & z err \\
\hline
\object{SDSS J000003.33-104315.8} & Sb(f)  & 17.32 &  0.09 & 16.57 &  0.04 & 16.01 &  0.06 & 22.30 & 21.56 & 21.00 & 0.0829 & 0.1038E-03  \\  
\object{SDSS J000151.29-092430.3} & Scd(f) & 17.71 &  0.03 & 17.13 &  0.02 & 16.89 &  0.04 & 22.45 & 21.86 & 21.63 & 0.0760 & 0.5085E-04  \\  
\object{SDSS J000222.39-102543.6} & Sd(f)  & 18.51 &  0.06 & 17.93 &  0.05 & 17.59 &  0.05 & 22.89 & 22.30 & 21.96 & 0.0000 & 0.0000E+00  \\  
\object{SDSS J000301.47-001901.8} & Sb(f)  & 17.51 &  0.13 & 16.60 &  0.10 & 16.13 &  0.10 & 21.60 & 20.69 & 20.23 & 0.0843 & 0.9440E-04  \\  
\object{SDSS J000347.01-000350.3} & Sc(f)  & 17.49 &  0.02 & 17.02 &  0.02 & 16.68 &  0.02 & 22.41 & 21.95 & 21.60 & 0.0625 & 0.6732E-04  \\  
\object{SDSS J000530.43-095701.2} & Sd(f)  & 17.67 &  0.04 & 17.20 &  0.03 & 16.86 &  0.04 & 22.74 & 22.27 & 21.93 & 0.0550 & 0.7188E-04  \\  
\object{SDSS J000542.53-111048.8} & Sb(f)  & 17.40 &  0.04 & 17.10 &  0.05 & 16.91 &  0.04 & 22.31 & 22.01 & 21.82 & 0.0400 & 0.4851E-04  \\  
\object{SDSS J000619.88-092744.8} & Sc(f)  & 16.93 &  0.01 & 16.29 &  0.02 & 15.95 &  0.01 & 21.85 & 21.21 & 20.88 & 0.0555 & 0.1026E-03  \\  
\object{SDSS J000628.86-004702.9} & Sa(f)  & 16.08 &  0.03 & 15.16 &  0.03 & 14.71 &  0.03 & 19.66 & 18.75 & 18.29 & 0.0443 & 0.9348E-04  \\  
\object{SDSS J000641.10-105825.9} & Sc(f)  & 16.64 &  0.84 & 16.25 &  0.99 & 15.90 &  0.87 & 22.89 & 22.50 & 22.15 & 0.0227 & 0.5682E-04  \\  
\object{SDSS J000741.22-004145.3} & Sc(f)  & 16.92 &  0.02 & 16.18 &  0.02 & 15.81 &  0.02 & 21.42 & 20.67 & 20.30 & 0.0735 & 0.1082E-03  \\  
\object{SDSS J000919.54+003557.2} & Sb(f)  & 16.52 &  0.10 & 15.72 &  0.08 & 15.30 &  0.06 & 21.11 & 20.30 & 19.88 & 0.0600 & 0.9067E-04  \\  
\object{SDSS J000924.17+003216.2} & Scd(f) & 18.36 &  0.04 & 17.62 &  0.02 & 17.44 &  0.05 & 22.76 & 22.01 & 21.83 & 0.0798 & 0.6654E-04  \\  
\object{SDSS J000941.16-003152.4} & Sa(f)  & 16.42 &  0.04 & 15.69 &  0.04 & 15.31 &  0.04 & 20.85 & 20.12 & 19.73 & 0.0000 & 0.0000E+00  \\  
\object{SDSS J000949.38-004103.1} & Scd(f) & 17.94 &  0.04 & 17.39 &  0.03 & 17.04 &  0.04 & 22.70 & 22.15 & 21.80 & 0.0390 & 0.5695E-04  \\  
\hline
\end{tabular}
\end{sidewaystable*}

The main catalog is listed in Tables \ref{cat1} and \ref{cat2}.  The
structural parameters of the catalog entries are shown in Table \ref{cat1}. Table \ref{cat2}
contains the photometric parameters and the redshifts.  All entries are
ordered by increasing right ascension.  The full tables are available from
the CDS.  These tables contain all edge-on galaxies with disks that fulfill
our automatic selection criteria, ranging from early-type spirals to
late-type spirals and irregulars.  In total, our catalog contains 3169
objects.  

Table \ref{cat1} is organized as follows: Column (1) presents the galaxy
name in the SDSS nomenclature, which is consistent with the IAU
nomenclature requirements.  The following two columns (2) and (3) contain
the coordinates of the galaxies, i.e., right ascension and declination for
the epoch J2000.  Column (4) indicates the general class (simple disks:
Sd(f); intermediate types: Sc(f), Scd(f) and Irr(f); galaxies with bulge:
Sa(f) and Sb(f)). The columns (5), (6), (7) show our derived value of
$\varepsilon$ in the {\em g}, {\em r}, and {\em i} bands.  Columns (8),
(9), and (10) contain the CI in these same bands.  The angular diameter in
{\em g}, {\em r}, and {\em i} in arcsec is presented in columns (11), (12),
and (13).  The axial ratios in the {\em g}, {\em r}, {\em i} bands are
derived from the ratio of $isoA/isoB$ of the SDSS parameters and are listed
in columns (14), (15), and (16).  {\em isoA} is the isophotal major axis and {\em isoB}
the isophotal minor axis of an isophote with a surface brightness of $\mu =
25$ mag arcsec$^{-2}$ (in the respective band) as given by the DR1
pipeline.

Table \ref{cat2} starts with the SDSS designation of the galaxy
(column (1)) followed by our proposed class in column (2).  Petrosian
magnitudes and their uncertainties are provided in the {\em g, r, and
i} bands in columns (3) to (8).  The total surface brightness in {\em
g, r, i} is given in columns (9), (10), and (11). We derived it using
the parameters $petroMag+rho$, which are given in the SDSS DR1
database. In this database, {\em rho} is  five times the logarithm of
the Petrosian radius in the {\em i} band.  The Petrosian magnitudes
and their uncertainties as well as the total surface brightnesses were
adopted from the SDSS DR1 archive. The Petrosian magnitudes and the
total surface brightnesses are corrected for Galactic foreground
extinction with the extinction values of \citet{schle98} as given in
the SDSS database.  

If a value of the
spectroscopic redshift {\em z} and its uncertainty are available from the SDSS
database, this measurement is listed in columns (12) and (13); otherwise
these values are denoted as zero.

\section{Completeness considerations from sample comparisons}

\subsection{A comparison of our automatically selected galaxy sample with
our visually classified sample}

In order to estimate the completeness of our general morphological groups,
we compare the results from our code with those from the visual inspection
using the reference set that we compiled to automate the separation
process.  Here we compare the three general classes, which are galaxies
with bulge, intermediate types, and simple disk objects.  According to the
previously defined limiting values for the three general types, class Sd(f)
contains 501 simple disks.  The comparison with our visual galaxy
classification shows an agreement of 97\%.  Using our automated
procedure, we identified 1065 objects from the bulge class (Sa(f) and
Sb(f)). 88\% of these were also identified as clear bulge galaxies in our
visually selected reference set.   The intermediate class contains the
largest fraction of galaxies (1603 objects).  This class is not 
homogeneous and contains Sc(f), Scd(f) and Irr(f) types.  The
comparison with the reference set indicates a completeness of 69\%, i.e., 
31\% belong to other classes using visual inspection.  88\%
of these 31\% seem to belong to the simple disk class when classified by eye.
This suggests that the automated classified intermediate class contains a
large number of Sd(f) types.  

The division between the classes is necessarily somewhat arbitrary, and
galaxies close to a boundary may in many cases also be considered members
of the adjacent class.  For instance, uncertainties can be introduced by
variations in galaxy properties which are a form of ``cosmic noise''.  We
required our visual subdivision to be consistent with the earlier studies
\citep{kar93,kar}.

The number of Sd(f) objects is relatively low.  However, we intentionally
chose fairly conservative separation criteria in order to minimize possible
contamination of our thus selected simple disk sample.  If we use a more
generous lower limiting value of $\varepsilon$$\geqslant$ 0.75 and include
the Scd(f) objects as simple disks, the simple disk object class contains
1004 galaxies (extended simple disk sample of seemingly bulgeless types, Scd(f) and Sd(f)).  This
corresponds to 32\% of the total catalog and enlarges the simple disk class
by a factor of two.  With this selection, however, the contamination by
other types is larger than with the more rigorously defined limits for
simple disks.  Table \ref{tbl-2} contains the absolute numbers and
percentages of the various classes in comparison to the entries in the
catalog as a whole.

\begin{table}
\caption{Galaxy classes and their fractions. The absolute numbers of
galaxies in the various morphological subclasses and their percentages 
with respect to the catalog entries as a whole are listed in this table.}
\label{tbl-2}
\centering
\begin{tabular}{rrr}
\hline\hline
General Class & Number & Percentages\\
\hline
Sa(f) & 222 & 7.01 \\
Sb(f) & 843 & 26.60 \\
Sc(f) & 1005 & 31.71 \\
Scd(f) & 503 & 15.87 \\
Sd(f) & 501 & 15.81 \\
Irr(f) & 95 & 3.00 \\
Total & 3169 & 100.00 \\
\hline
\end{tabular}
\end{table}

\subsection{A comparison of the Revised Flat Galaxies Catalog with our
catalog}

We searched for the RFGC galaxies in the SDSS DR1 using the coordinates
given in the RFGC and recovered 328 objects.  Then we checked how many of
these galaxies are recovered in our catalog using our selection criteria.
We found 273 objects in common.  

The remaining 55 RFGC galaxies were studied to find out why they were not
recovered. In most cases objects are not recovered because they are not
detected as SDSS targets (``Photoobjects''). This is the case when a galaxy
is located near the borders of an SDSS stripe, which has the consequence
that this object is not included in the ``Best Galaxy Table'' and
subsequently not detected in the SDSS ``Galaxy'' catalog. In the cases of
relatively extended objects these systems are ``shredded'' by the SDSS
detection software and thus not included in our catalog. Furthermore, there
are a few cases where the SDSS shows a galaxy with an inclination deviating
from an edge-on orientation at a given set of RFGC coordinates.  A small
subset of RFGC galaxies are not really edge-on galaxies.  If RFGC galaxies
are very close to nearby bright foreground stars, they are also rejected by
the SDSS software.  We conclude that we recovered all RFGC galaxies that
conform to our selection criteria except for those missed by the SDSS
software and for those that are not edge-on.  Hence the RFGC is more
complete than our catalog for nearby (and hence seemingly large) edge-on
systems.  The RFGC is also more complete in terms of spatial coverage since
it does not suffer from the detection problems near the bright stars or
edges of stripes.

\begin{figure*}
\resizebox{\hsize}{!}{\includegraphics[angle=0,width=\textwidth,]{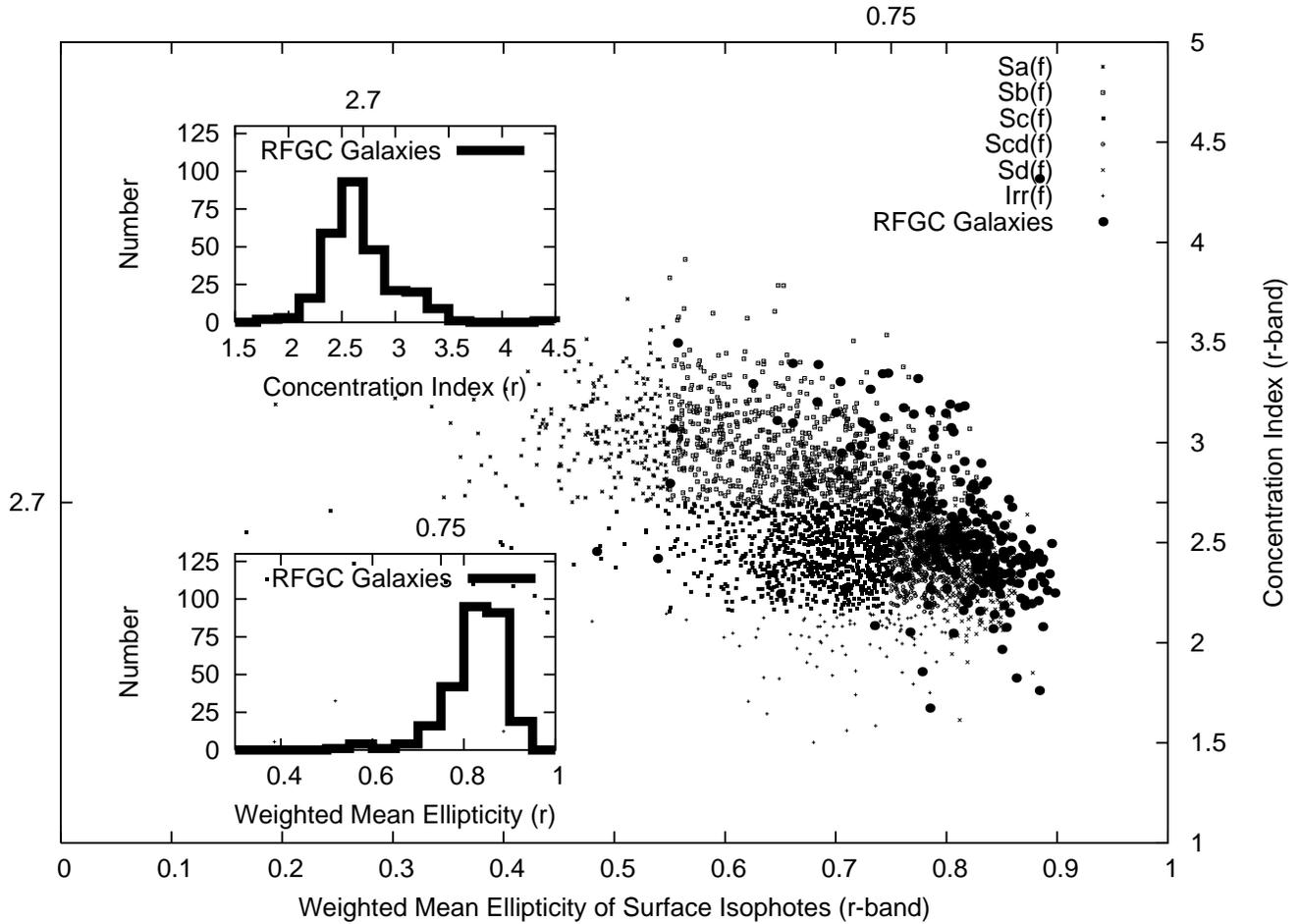}}
\caption{Separation diagram with the recovered RFGC galaxies. Galaxies of
the different morphological types are indicated by small symbols as given
in the legend, recovered galaxies from the RFGC with large filled points.
The upper inset shows the number distribution for the recovered RFGC
galaxies of the concentration index, the bottom inset that of the weighted
mean ellipticity.}
\label{sdrfgc}
\end{figure*}

We plot the location of the recovered RFGC galaxies in our separation
diagram in Fig. \ref{sdrfgc}.  It is clearly seen that most of the RFGC
objects belong to the Scd(f) and Sd(f) class (184 of 273).  Additionally, a
smaller number of RFGC systems is found in the Sb(f), Sc(f) and Irr(f)
classes.  This is illustrated in the two inserted histograms in Figure
\ref{sdrfgc}: RFGC objects clearly exceed our chosen limits for simple disk
galaxies in the case of both discriminators, the CI (upper inset) and the
$\varepsilon$ (bottom inset).  That means that \citet{kar93} and
\citet{kar} did not only select simple disk systems, but also included some
mixed-morphology galaxies, whose classification we can now correct thanks
to the CCD data.{\footnote{Note, however, that our catalog contains {\em
all} edge-on disk galaxies that we identified with our selection criteria,
including mixed-morphology and bulge-dominated galaxies.  The galaxy type
can be found in Table \ref{cat1} and Table \ref{cat2} as explained earlier.}

With our separation routine we have thus improved the identification of
simple disks in contrast to the flat galaxies catalogs.  There are several
reasons for this improvement: One is that the CCD images of the SDSS have a
greater uniformity than the earlier employed blue POSS-I and ESO/SERC photo
plates.  In addition, the SDSS imaging data are deeper -- they are 50\%
complete for point sources at g = 23.2 mag \citep{aba03}. The limiting
magnitude of the blue POSS-I plates is 20 mag(R) \citep[see][page 481]{min63}, 
and of the ESO/SERC J plates is 22.5 mag(B) \citep{reid91}.  Because of the
higher sensitivity, depth, and resolution of the SDSS, we can identify more
substructure within our galaxies, which leads to a more accurate
classification. This improved classification benefits from our choice of
the SDSS r-band, which has the highest photon efficiency in the SDSS.

Our selection parameters include galaxies with smaller diameters and extend
to fainter magnitudes.  A comparison of the magnitudes and diameters of the
recovered RFGC objects and the remaining galaxies from our catalog is shown
in Fig.\ \ref{diarfgc}.  As is to be expected, the figure shows that our
catalog contains objects with smaller diameter as well as fainter objects.
Furthermore, our catalog contains galaxies not detected in the RFGC but
with diameters and luminosities in the range of the RFGC galaxies.  Note
that we also have a larger number of more luminous galaxies at a given
diameter.  This is the consequence of permitting smaller axial ratios than
the RFGC, which favors a larger number of disk galaxies with a bulge.
Moreover, we show in Fig.\ \ref{abrfgc}, that in order to recover the RFGC
galaxies from the SDSS database, we need to use selection criteria with
smaller axis ratios.  The SDSS axis ratios tend to be smaller than those of
the RFGC.  When comparing the number of seemingly bulgeless types (our Scd(f) and Sd(f)
class with $\varepsilon \geqslant$ 0.75 and CI $<$ 2.7), our catalog
contains 1004 objects including 184 RFGC galaxies (out of 273 RFGC galaxies
within the DR1 area).  While we are missing galaxies near the edges of
stripes etc., we still have $\sim 3.7$ times more simple disk galaxies than
were found in the RFGC within the same area.  As Fig.\ \ref{diarfgc}
illustrates, this is only in part because of the inclusion of smaller axis
ratio.  We attribute it also to the higher sensitivity and homogeneity of
the SDSS.  

\begin{figure}
\resizebox{\hsize}{!}{\includegraphics[angle=0,width=\textwidth,]{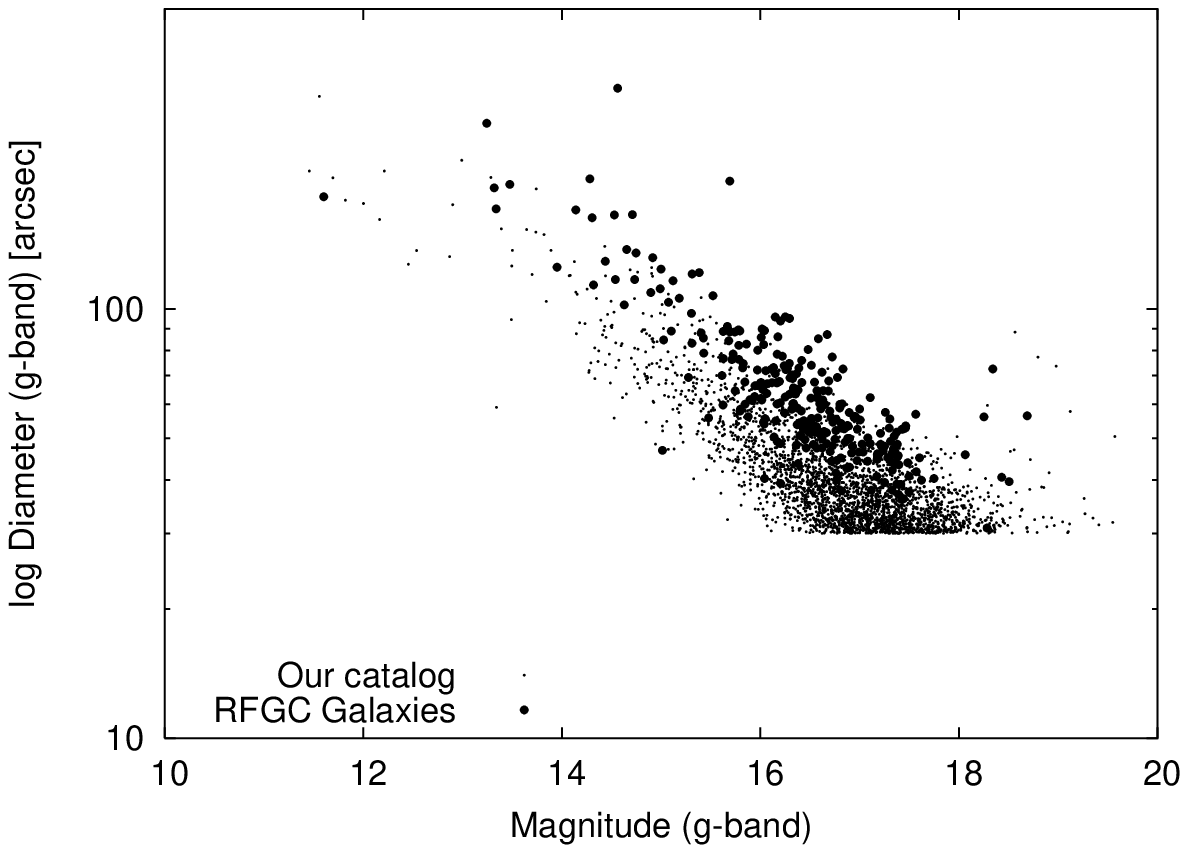}}
\caption{Comparison of the depth of this catalog versus the recovered RFGC 
objects. The objects of the RFGC are large filled points, the catalog 
galaxies are small dots.}
\label{diarfgc}
\end{figure}
It is difficult to determine absolute completeness numbers for
our survey.  For instance, in order to recover the initial RFGC training
set, we had to decrease the angular diameters as compared to the parameters
chosen in the RFGC.  Furthermore, some incompleteness effects will affect
all galaxies alike (e.g., the non-detection due to the location close to
the border of a stripe), whereas for instance dust will affect certain
galaxy types in particular (see next section). 

\begin{figure}
\resizebox{\hsize}{!}{\includegraphics[angle=0,width=\textwidth,]{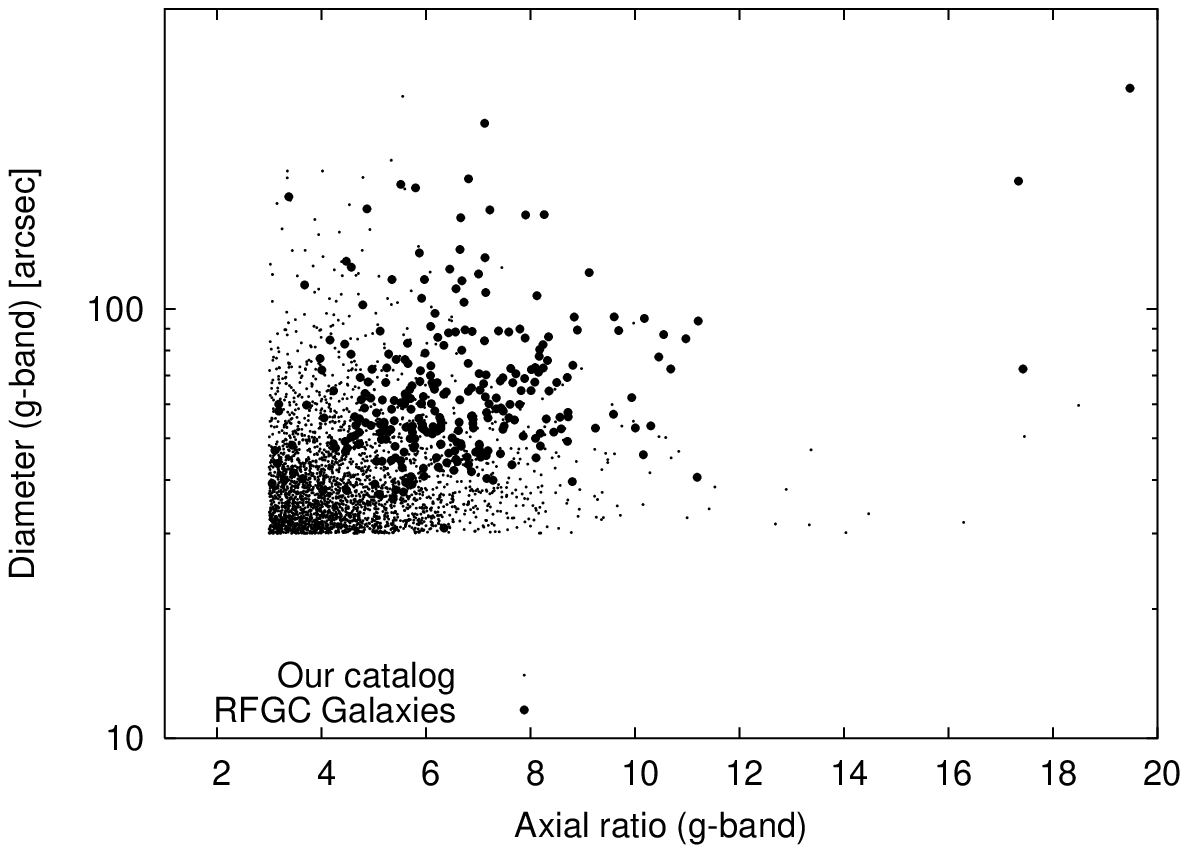}}
\caption{Comparison of the depth of this catalog versus the recovered 
RFGC objects. The objects of the RFGC are large
filled points, the catalog galaxies are small dots.}
\label{abrfgc}
\end{figure}

\section{The influence of dust and distance} \label{dust}

We have subdivided our edge-on galaxies in objects with bulge and in simple
disk systems without a bulge component. In this section we discuss the
expected influence of dust on our separation procedure and a resolution
bias caused by distance.

The distribution of dust in spiral galaxies has been the subject of a
lively debate over decades.  Recent studies try to model the influence of
dust on the surface photometry. \citet{kuc98} compared the optical/near
infrared (NIR) color gradients of edge-on galaxies with the reddening from
radiative transfer models. These models use Monte-Carlo techniques in order
to describe the radiative transfer of photons (including scattering,
absorption, and re-emission) in different dust environments \citep{gor01}.
These models were then adapted to edge-on galaxy examples \citep[see
e.g.,][]{xil97,xil98,poh00,pop00,mis02}.  The best models include a
homogeneous and clumpy distribution of dust \citep{kuc98}. \citet{tuf04}
computed the attenuation of stellar light at different inclinations,
wavelengths, and opacities from the different geometrical components of a
spiral galaxy. They found that the extinction strongly depends on the
inclination. In the case of edge-on systems most of the attenuation by dust
occurs in the thin disk component, which often includes a typical dust
lane.

But the amount of dust in different edge-on spiral types is not constant.
This was recently shown by \citet{ste04} with new SCUBA observations. Their
measurements show that the flat galaxy \object{NGC 5907} (\object{FGC 1875}) 
contains a very
high amount of neutral hydrogen but only small amount of total dust.  A
high ratio of the mass of the neutral hydrogen to the mass of cold dust
implies a very low star formation efficiency. In addition, \citet{mat99}
and \citet{mat00} show the ``lack of a quintessential dust lane'' in the
prototypical superthin galaxy \object{UGC 7321}. In another paper, \citet{mat01}
conclude that the dusty interstellar medium (ISM) in this type of objects
has a clumpy and patchy distribution. They derived the observed properties
of dust with the aid of a multiphase ISM model and found that
``$\approx$50\% of the dusty material in \object{UGC 7321} is contained in a clumpy
medium''. The other half has a diffuse distribution. UGC 7321 is an LSB
galaxy with a large axial ratio and no bulge component, a typical simple
disk.

What are the differences in the  properties of edge-on galaxies with
an organized dust lane and those that exhibit a clumpy and diffuse
dust distribution? \citet{dal04} found a clear boundary between
edge-ons with and without a dust lane.  They conclude that the dust
distribution is connected with the rotation velocity, i.e., galaxy
mass.  Organized dust lanes appear in high surface brightness objects
with a relative rapid rotation velocity. In galaxies with rotation
velocities below $V_c = 120$ km s$^{-1}$ the dust has not settled into
a thin dust lane.  The dust distribution of these simple disk galaxies
with typically a low surface brightness is clumpy and diffuse out to
large scale heights. 

What is the effect of a clumpy or an organized dust distribution on our
separation values?  In galaxies with a small angular size the dust lanes
are unresolved, especially in those with larger distances. For these
galaxies it is not a simple exercise to verify the existence of an
intrinsic dust lane. In some of our larger and brighter galaxies a dust
lane is visible, in others not.  The dust is concentrated in the thin disk
component and consequently the light attenuation caused by the thin disk is
the highest \citep{tuf04}.  The stellar disk disappears in the extreme
cases where a large amount of dust extends beyond the whole disk.
Therefore the consequences of the dust attenuation of the disk are expected
to be stronger in more massive galaxies like \object{NGC 891}. In these objects the
probability of a dominant bulge is high.  They are more metal-rich and are
expected to contain more dust \citep[mass-metallicity relation,
e.g.,][]{dal04}.  In general, the dust dims the thin disk with respect to
the bulge brightness. This circumstance introduces three additional biases
for our catalog:

\begin{enumerate} 
\item Early-type disk galaxies: Our catalog has an incompleteness for
galaxies with big bulges and attenuated underlying thin disks. In these
cases the disk cannot easily be detected as a bright structure because its
morphology is like a dark line in a bright bulge component.  Therefore this
catalog is likely to be incomplete for dusty early-type spirals, especially
for the types S0(f) and Sa(f), where we may be missing the disks.  This may
lead to an overestimate of the number of simple disks as compared to
pronounced bulge-disk systems.  
\item Massive late-type spirals/massive simple disks: A bias is shown in
the extreme cases where the thin disk of a high-mass simple disk galaxy is
almost completely obscured by dust.  In this case the galaxies may have too
small an axial ratio $a/b$ to be selected for our catalog.  In the less
extreme cases of high-mass simple disk galaxies the dust dims the thin
disk. The influence of dust in a simple disk is that the value of
$\varepsilon$ decreases because a dusty disk appears to be thicker. For
that reason, the affected galaxies may exhibit an offset toward the
intermediate class in the separation diagram (Fig.\ \ref{sd}). The CI of
such galaxies is not distorted because these objects have no distinct
bulge.  
\item Objects at large distances are affected by a resolution bias. The
consequences are that an unresolved dustlane dims the light of the stellar
disk compared to a dust-free disk. This does not affect $\varepsilon$; the
disk looks simply smaller. This may lead to a slight increase of the CI for
disks with a bulge. However, the separation of the general classes does not
seem to be displaced by the presence or absence of unresolved dust lanes.
The important effect is that a strongly dimmed disk looks smaller and may
fail to pass the selection criterion.  The catalog will be more incomplete
for small angular size disk galaxies with unresolved dust lanes.  To reduce
this effect we impose a minimum diameter in our search.  
\end{enumerate}

A future paper is planned in order to explore the effects of dust using
simulated galaxies with varying amounts of dust and inclinations.

\section{Discussion} \label{Di}

We identified edge-on disk galaxies in the SDSS DR1, which we subdivide in
the following classes:

\begin{description}
\item
{\bf Disk galaxies with a bulge; Sa(f), Sb(f)} (CI $\geqslant$ 2.7): 
The fraction of these objects in the catalog is 34\%. This
class contains spiral galaxies with a bulge that are not affected by the
selection effect described in section \ref{dust}.
\item
{\bf Intermediate class; Sc(f), Scd(f), Irr(f)} (CI $<$ 2.7 \& $\varepsilon
<$ 0.8): These late-type galaxies show a central light concentration and
often a bouffant disk but no obvious bulge.  
With a fraction of 50\%
these types represent the majority in this catalog.
These galaxies may have an
inclination slightly smaller than edge-on or may show pronounced warps.  At
CI $<$ 2.15 \& $\varepsilon <$ 0.8 the class of disky edge-on (dwarf)
irregulars appears.  They show an asymmetric ``puffy'' disk with small
clumpy (not central) light concentrations comparable to those found by
\citet{par02} or blue compact dwarfs \citep[e.g.,][]{san84}; see also
\citet{kniazev04a}.
\item
{\bf Simple disk galaxies; Sd(f)} (CI $<$ 2.7 \& $\varepsilon \geqslant$
0.8): These galaxies appear to be pure bulgeless disks. 
Using the conservative separation values, this class contains 16\%
of the
catalog objects. 
\end{description}

In order to check the usefulness of our separation we visually inspected
galaxies in the extreme regions in our separation diagram and found the
following subgroups:

\begin{description}
\item 
{\bf Dusty disk-dominated galaxies} (CI $\gtrsim$ 2.6 \& $\varepsilon
\gtrsim$ 0.75): These types have flat extended disks and slight central
light concentration. The majority of this type appears as extended disks
with dust lanes, small bulges and very blue outer disks.
\item
{\bf Complex bulge/disk systems} (CI $\gtrsim$ 3 \& $\varepsilon <$ 0.7):
These types are mostly complex bulge-disk systems. The bulge of these
systems becomes clearly visible. In some of these types a stellar disk
extends out to large scale heights and forms a bright but diffuse envelope
around the galaxy.
\end{description}

In addition to the visual inspection of all galaxies in the extreme regions
in the $\varepsilon$--CI space we also checked galaxies located in the
central regions of the selection boxes for every general type by eye.  In
agreement with our expectations we found simple disk systems at high
$\varepsilon$ and low CI (0.8 $<\varepsilon<$ 0.85 \&  2.3 $<$CI$<$ 2.4).
Their appearance is blue and needle-like. Intermediate values of CI and
$\varepsilon$ reveal the region where intermediate types are concentrated
(0.7 $<\varepsilon<$ 0.75 \& 2.5 $<$CI$<$ 2.6).  This group is dominated by
lenticular-shaped puffy disks and smooth central light concentration but no
dominant bulge component. The center is slightly redder than the bluer
outer parts in these systems. It seems that they often have extended faint
LSB disks around the bright parts.  We checked the central region of disk
galaxies with dominant bulges (0.65 $<\varepsilon<$ 0.7 \& 3 $<$CI$<$ 3.1).
These early-type spirals are visibly less blue than the other general types
in this catalog, and galaxies with bulge are the less well-populated group.

The highest concentration of galaxies can be found in the transition zone
between the intermediate types and the simple disks. This indicates the
lack of clear-cut boundaries between different types; instead we are seeing
a continuum.  

Fig. \ref{mu} shows a comparison of the
general classes with surface brightness. As shown in this
Figure, the presence or absence of bulges has an influence on the
overall
surface brightness of a galaxy as one would expect.  Simple
disk galaxies have the lowest intrinsic surface brightnesses of all
edge-on galaxies in this catalog. 
For this plot we use the surface brightness given in column 10 of the Table \ref{cat2}.
This surface brightness is derived as explained in Section \ref{os}.
No dust correction is applied.

\begin{figure*}
\resizebox{\hsize}{!}{\includegraphics[angle=0,width=\textwidth,]{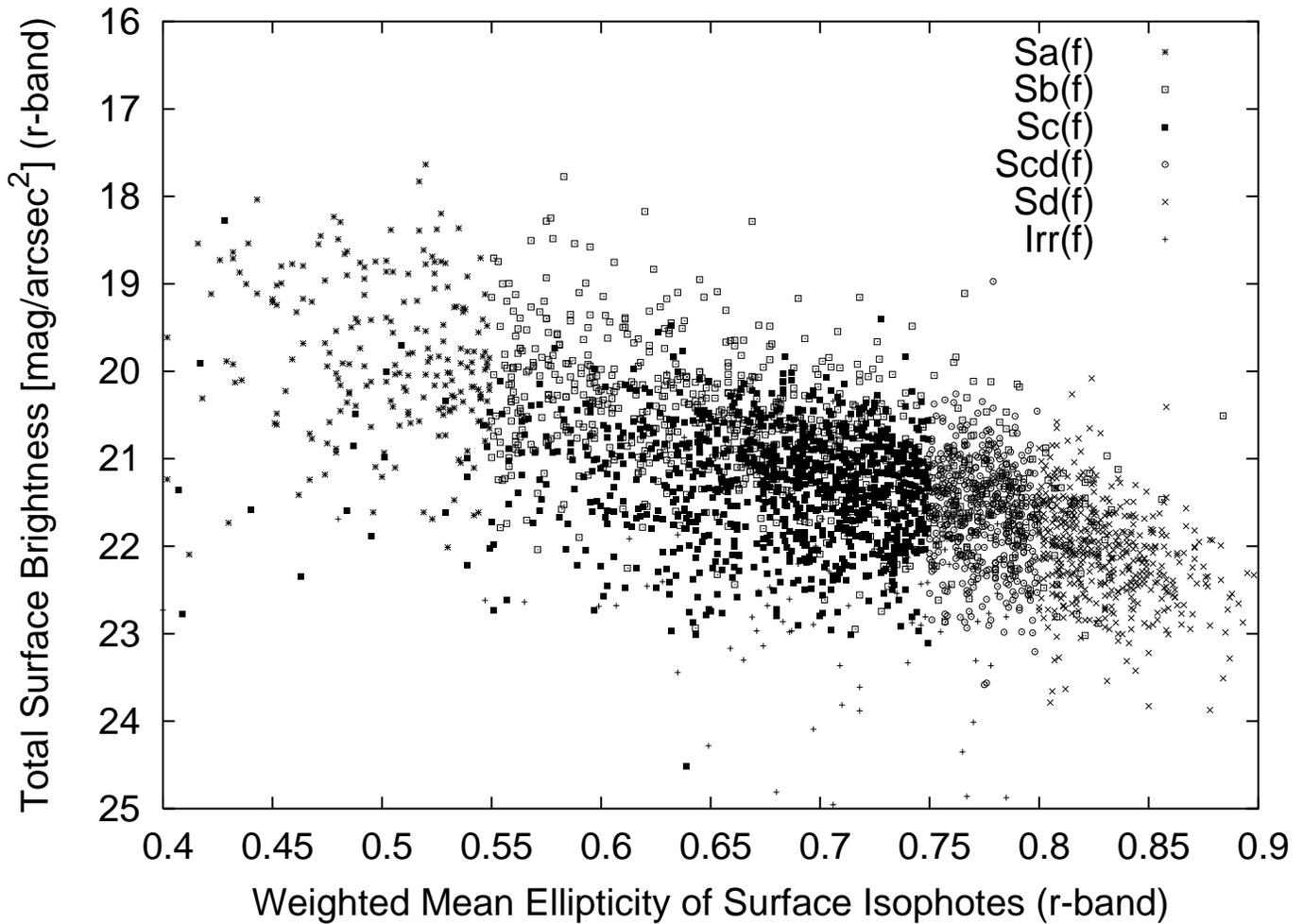}}
\caption{Distribution of the $\varepsilon$ versus the total surface brightness. The different points represent the various
morphological galaxy types and are explained in the key.}
\label{mu}
\end{figure*}

\section{Conclusions \& Summary} \label{CS}

We present for the first time a homogeneous and large dataset of uniformly
selected edge-on disk galaxies.  These common galaxy types are very
important in order to understand the formation and evolution of disk
galaxies. The galaxies are selected from the SDSS DR1 
on the basis of photometric structural parameters indicative of extended stellar
disks with large major to minor axis ratios.
 
With the aid of this method we gathered 3169 edge-on galaxies in an
area of 2099 deg$^2$ and selected through an automated separation
algorithm. Using visual inspection of the galaxies we realized that this
catalog contains three general classes of crude morphological types: disk
galaxies with bulge, thin bulgeless objects and intermediate types. 

The separation is based on the central light concentration and the flatness
of the galaxy images.  The light concentration is expressed via the
concentration index CI. With this CI we are able to distinguish between
disk galaxies with a bulge component Sa(f) and Sb(f) and those with
bulgeless appearance -- the Sc(f), Scd(f), Sd(f) and Irr(f) classes. As a
second discriminator we use the luminosity-weighted average of the
ellipticity $\varepsilon$, derived from elliptical isophotes of every
object.  The $\varepsilon$ allows one to distinguish 
several structural groups with flat disks: 
the early-type spirals, types 
Sa(f) from Sb(f); the apparently bulgeless systems from intermediate types
Sc(f), Scd(f) and Irr(f) with central light concentrations; and the thin,
smooth galaxies, the simple disk class Sd(f). 

The simple disk class includes objects previously defined as flat and
superthin galaxies and it exhibits the lowest surface brightness compared
to the other classes. The axial ratios of simple disk galaxies are the
largest. The intermediate class of edge-ons is composed of different types
of galaxies including (dwarf) irregular systems. There is no well-defined
boundary between these general classes, but instead a continuum of
properties exists.

The fraction of galaxies with bulge (Sa(f), Sb(f)) is 34\%, those without
bulge 16\% (Sd(f)) and the fraction of the intermediate class is 50\%.
However, we found that the intermediate object class contains also a large
fraction (about 440 objects, these are nearly 30\% of the intermediates) 
of seemingly bulgeless types.  Therefore we conclude that
every general class (i.e., galaxies with bulge, intermediate objects and
simple disks) represents about one third of the galaxies listed in the
catalog. The true numbers of our classes are somewhat lower.  Dust
attenuation introduces a bias such that this catalog is expected to be
incomplete for early-type spirals with pronounced dust lanes.  Additional
incompleteness is introduced because of features of the SDSS database such
as galaxy shredding, etc. 

In the case of late-type spirals, dust is expected in high-mass systems and
increases the apparent thickness of the disk.  This results in a minor
offset in the separation but has no effect on the completeness for simple
disks.  Unresolved dust lanes dim the disk light and lead to higher
incompleteness for distant disk galaxies.  A comparison with the RFGC shows
that our catalog suffers from incompleteness for, e.g., galaxies close to
bright stars or near the edge of a scan stripe, but that it nonetheless
contains almost four times as many galaxies within a given area than the
RFGC.  This is mainly because we included also galaxies with smaller
angular diameters, but it is also a result of the homogeneity, resolution,
and depth of the SDSS.

This catalog provides a large, homogeneously selected galaxy sample for
which sensitive five-color photometry (and in many cases also spectroscopy)
is available.  SDSS spectroscopy, while covering only a portion of the
galaxies because of its circular aperture of $3''$, will be useful for a
wide variety of studies, for instance for deriving metallicities and for
constraining the properties of the underlying stellar populations
\citep[e.g., ][]{kniazev04b,bernardi05}. 

While these data will be analyzed in later papers, even the raw catalog
data have interesting implications.  Our results re-enforce the conclusions
of Karachentsev and collaborators \citep{kar93,kar} that simple disk
galaxies are relatively common, especially among intermediate-mass
star-forming galaxies \citep{matt97}.  Galaxy formation models must be able
to produce such high angular momentum systems with reasonable frequencies.
We also find that the simple disks are not a separate morphological class,
but rather are at the end of a continuum that extends smoothly from
bulge$+$disk systems. However, the simple disks tend to be lower surface
brightness galaxies, indicating that the probability for bulge formation
depends on host galaxy mass. This in turn can be linked to models where
bulges form from internal disk instabilities through the dependence of the
Toomre Q-parameter on disk surface density \citep[e.g.,][]{Immel04}.
Similarly the properties of our sample will be useful in constraining the
role of galaxy mergers in building disk-halo galaxies \citep[e.g.,][]{spri05,korm04}. 
We will explore these and related issues in future papers.

\begin{acknowledgements}
SJK and EKG were supported by the Swiss National Science Foundation through
the grants 200021-101924/1 and 200020-105260/1.  JSG thanks the U.S.
National Science Foundation for support through grant AST-9803018 to the
University of Wisconsin, and the University's Graduate School for
additional funding of this work. We thank our referee, Dr.\ Igor
Karachentsev, for his thoughtful comments.

Funding for the creation and distribution of the SDSS Archive has been
provided by the Alfred P. Sloan Foundation, the Participating Institutions,
the National Aeronautics and Space Administration, the National Science
Foundation, the U.S. Department of Energy, the Japanese Monbukagakusho, and
the Max Planck Society. The SDSS Web site is http://www.sdss.org/.

The SDSS is managed by the Astrophysical Research Consortium (ARC) for the
Participating Institutions. The Participating Institutions are The
University of Chicago, Fermilab, the Institute for Advanced Study, the
Japan Participation Group, The Johns Hopkins University, the Korean
Scientist Group, Los Alamos National Laboratory, the Max-Planck Institute
for Astronomy (MPIA), the Max-Planck Institute for Astrophysics (MPA), New
Mexico State University, University of Pittsburgh, University of
Portsmouth, Princeton University, the United States Naval Observatory, and
the University of Washington.
\end{acknowledgements}

\end{document}